\documentclass[aps,prd,twocolumn,groupedaddress,nofootinbib]{revtex4-2}

\usepackage{dsfont}
\usepackage{natbib}
\usepackage{url}
\usepackage{color}
\usepackage{fullpage}
\usepackage[top=2cm, bottom=4.5cm, left=2.5cm, right=2.5cm]{geometry}
\usepackage{amsmath,amsthm,amsfonts,amssymb,amscd}
\usepackage{lastpage}
\usepackage{enumerate}
\usepackage[shortlabels]{enumitem}
\usepackage{fancyhdr}
\usepackage{mathrsfs}
\usepackage{xcolor}
\usepackage{listings}
\usepackage{dcolumn} % Align table columns on decimal point
\usepackage{bm}
\usepackage{hyperref}
\usepackage{graphicx}
\hypersetup{%
  colorlinks=true,
  linkcolor=blue,
  linkbordercolor={0 0 1}
}

\newcommand{\be}{\begin{equation}}
\newcommand{\ee}{\end{equation}}
\newcommand{\bea}{\begin{eqnarray}}
\newcommand{\eea}{\end{eqnarray}}
\newcommand{\nn}{\nonumber}

\newcommand{\pp}{p_{\phi}}
\newcommand{\sq}{\sqrt{q}}
\renewcommand{\H}{\mathcal{H}}
\newcommand{\lv}{\lambda_v}
\newcommand{\lp}{\lambda_{\phi}}
\newcommand{\sv}{\sigma_v}
\renewcommand{\sp}{\sigma_{\phi}}

\begin{document}

\title{Polymer cosmology with polymer matter: Effective dynamics}

\author{Aleena Zulfiqar}%
 \email{aleena.706@lums.edu.pk}
\affiliation{%
 Department of Mathematics, Syed Babar Ali School of Science and Engineering, Lahore University of Management Sciences, Lahore 54792, Pakistan
}%
\author{Syed Moeez Hassan}
 \email{syed\_hassan@lums.edu.pk}
 \affiliation{%
 Department of Physics, Syed Babar Ali School of Science and Engineering, Lahore University of Management Sciences, Lahore 54792, Pakistan
}%

\date{\today}

\begin{abstract}

We study the effective dynamics of a polymer quantized scalar field living on an effective polymer quantized homogeneous and isotropic background. We use a presureless dust field as an internal clock, and work with volume variables. Our results show how the quantum bounce in the early universe is affected as the initial conditions and various physical parameters - including the scalar field polymer scale - are varied. We also find, generically, that there is asymmetric evolution across the bounce.

\end{abstract}

\maketitle
%\tableofcontents

\section{\label{sec-intro} Introduction}

The construction of a quantum theory of gravity is an open problem with multiple contending theories, each with its own strengths. In the quantization of pure gravity, progress has been made in the Loop Quantum Gravity (LQG) program where a kinematical Hilbert space (along with suitable operators on it) is available. Loop Quantum Cosmology (LQC) is the application of techniques from LQG to homogeneous cosmologies where things are much simpler. One main technique that has emerged from these programs is the idea of ``polymer quantization'' (which is distinct from standard `Schr\"{o}dinger quantization'), where a fundamental discreteness (a polymer scale) is built into the theory at the quantum level \cite{Ashtekar:2002sn, Halvorson_2004}.

Both LQC models \cite{Ashtekar:2011ni}, and polymer quantized cosmologies \cite{Husain:2003ry, BenAchour:2018jwq, Montani:2018uay, Barca:2019ane, Giovannetti:2019ewe, Giovannetti:2020nte, Giovannetti:2021bqh, Barca:2024qls} have been studied in detail in the literature, with the main lesson that the initial `big bang' singularity is replaced by a quantum bounce (see \cite{Barca:2021qdn} for a review of and comparison between the two approaches). Some studies that are close to the model we present here include \cite{Pawlowski:2011zf} where the standard and loop quantizations of a homogeneous and isotropic Friedmann-Lemaitre-Robertson-Walker (FLRW) universe with a cosmological constant, and a massless scalar field (playing the role of an internal clock) have been investigated in detail; \cite{Husain:2011tm} where instead of the scalar field, a dust field is used as an internal clock; and a recent work \cite{Giesel:2020raf}, that investigates the role of different clock choices (including both the scalar field and the dust field) in LQC (with an inflationary potential) vis-\`a-vis singularity resolution and inflationary dynamics. However, all these models are either purely gravitational, or if there is a matter field included, it is either used as a reference clock, or its dynamics are treated via standard quantization methods.

An interesting question to ask is what happens if matter fields are polymer quantized? This question has been addressed for simple matter systems where differences appear in the energy spectrum and eigenfunctions \cite{Corichi:2007tf, Husain:2007bj}, for scalar fields in flat spacetime \cite{Ashtekar:2002vh, Hossain:2009vd, Husain:2010gb, Laddha:2010hp, Hossain:2010eb}, and for scalar fields in cosmology, where polymer quantization effects produce eras of inflation \cite{Hossain:2009ru, Hassan:2014sja, Hassan:2017cje, Ali:2017fhp} (an alternate polymer quantization scheme has been implemented in \cite{Kreienbuehl:2013toa} with cosmology coupled to a massless scalar field, where the scale factor plays the role of an internal clock, and the scalar field is polymer quantized. Their results show singularity resolution). In the cosmological case, however, the gravitational variables are treated classically, and the effective dynamics of polymer quantized matter is studied on this classical background.

In short, we either have polymer gravity with standard matter, or standard gravity with polymer matter (see \cite{Domagala:2012tq} for a rare exception where both sectors are polymer quantized, and also \cite{Laddha:2008em} where matter and `gravity-like' variables are polymer quantized). From a fundamental perspective, if polymer quantization is indeed the preferred method of quantization chosen by Nature, then it is natural to expect that both the gravitational and the matter sectors are polymer quantized. In this study, we present such a model for the homogeneous and isotropic FLRW universe with a cosmological constant minimally coupled to a scalar field and a pressureless dust field. The dust field is chosen as a matter reference clock, both the scalar field and the gravitational variables are polymer quantized, and we study the effective dynamics of the resultant theory.

Effective dynamics are obtained after taking the expectation values of the quantum Hamiltonian operator in some suitable `semi-classical' states \cite{Husain:2006uh, Taveras:2008ke}. These states have the property that they are sharply peaked on the corresponding classical phase space variables, and they achieve the minimum uncertainty bound for the canonically conjugate quantum operators. The resultant theory is classical, with quantum effects incorporated that become significant near the Planck scale. While the theory is not fully quantum, it reproduces to a fair bit of accuracy the corresponding quantum dynamics \cite{Rovelli:2013zaa}. Furthermore, the study of effective dynamics of polymer quantized theories has led to the prediction of many new phenomena \cite{Hossain:2009ru, Hassan:2014sja}, and serves as a useful guide on what to expect from the full quantum theory. A detailed analysis of the quantum bounce and effective dynamics of LQC (in particular, the dependence on state width), with a massless scalar field acting as a clock, and without a cosmological constant, has been carried out in \cite{Diener:2014mia}. A related analysis of the quantum bounce with respect to non-Gaussian initial states appears in \cite{Diener:2014hba}.

Our main results are that in the purely gravitational sector with dust time (without the scalar field), the location of the bounce depends on the gravitational polymer scale, the width of the gravitational semi-classical state, the cosmological constant, and also changes if the initial conditions are changed. In the theory with the polymer quantized scalar field included, we find that the dynamics are sensitive to the value of the matter (scalar field) polymer scale (which in general is distinct from the gravitational polymer scale), and in particular the location of the bounce changes as the matter polymer scale is varied. Furthermore, the evolution across the bounce is asymmetric in general. These results show how the quantum bounce is affected as various physical parameters are changed, and also illustrate the role of the matter polymer scale in the dynamics of the universe.

This paper is organized as follows. In the next section (\ref{sec-model}), we formulate our model, provide a brief review of polymer quantization, and derive the effective Hamiltonian. In Sec. (\ref{sec-dyn}), we discuss the dynamics of this theory and show numerical evolutions for both the pure gravity case, and for the gravity with scalar field case. We summarize and conclude in Sec. (\ref{sec-con}). We work in $c = \hbar = 8 \pi G = 1$ units throughout.

\section{\label{sec-model} The model}

We start with General Relativity written in the Arnowitt-Deser-Misner (ADM) form, coupled to a pressureless dust field $T$, and a scalar field $\phi$,
\be
\label{full-theory}
S = \int d^3x dt ~ [ \pi^{ab} \dot{q}_{ab} + \pp \dot{\phi} + p_T \dot{T} - N \H - N^a C_a],
\ee
where,
\bea
\label{full-ham}
\H &=& \H_G + \H_{\phi} + \H_D, \nn \\
\H_G &=& \dfrac{1}{\sq} \Bigg(\pi^{ab} \pi_{ab} - \frac{1}{2} \pi^2 \Bigg) + \sq ( \Lambda - R), \nn \\
\H_{\phi} &=& \dfrac{\pp^2}{2 \sq} + \dfrac{1}{2} \sq q^{ab} \partial_a \phi \partial_b \phi + \sq V(\phi), \nn \\
\H_D &=& p_T \sqrt{1 + q^{ab} \partial_a T \partial_b T }, \\
C &=& C^G_a + C^{\phi}_a + C^D_a, \nn \\
C^{G}_a &=& D_b \pi^b_a, \nn \\
C^{\phi}_a &=& \pp \partial_a \phi, \nn \\
C_a^D &=& -p_T \partial_a T,
\eea
$(q_{ab}, \pi^{ab})$, $(\phi, \pp)$ and $(T, p_T)$ are the gravitational, scalar field and dust phase space variables respectively, $q$ is the determinant of the spatial metric $q_{ab}$, $\pi$ is the trace of $\pi^{ab}$, $R$ is the 3-Ricci scalar curvature, $\Lambda$ is the cosmological constant, $V(\phi)$ is the scalar field potential, $N$ is the lapse and $N^a$ is the shift.

We then reduce to a spatially flat ($k=0$) FLRW universe with the ansatz,
\be
q_{ab} = v^{2/3}(t) e_{ab} , ~ \pi^{ab} = \dfrac{1}{2} v^{1/3}(t) p_v(t) e^{ab},
\ee
where $v(t)$ is the volume of the universe (related to the usual scale factor $a(t)$ as: $v = a^3$), $p_v$ is the momentum canonically conjugate to the volume, and $e_{ab} = \text{diag}(1,1,1)$ is the flat metric.

We also assume that the scalar and dust fields are homogeneous (to ensure consistency with an FLRW universe),
\be
\phi = \phi(t) ~ , ~ \pp = \pp(t),
\ee
\be
T = T(t), ~ p_T = p_T(t).
\ee
With this homogeneity ansatz, the diffeomorphism constraint vanishes identically and we are left with,
\be
S = V_0 \int dt ~ \Bigg( p_v \dot{v} + \pp \dot{\phi} + p_T \dot{T} - NH \Bigg),
\ee
where,
\be
\label{Ham-constr-cosm}
H =  -\dfrac{3}{4}v p_v^2  + v \Lambda  + \dfrac{\pp^2}{2 v} + v V(\phi) + p_T \approx 0
\ee
is the Hamiltonian constraint, and $ V_0 = \displaystyle \int ~ d^3x $ is the fiducial volume. The FLRW spacetime has a scaling symmetry as the line element is invariant under transformations of the form: $x \rightarrow lx, ~a \rightarrow a/l$ (where $l$ is some length scale). Exploiting this symmetry, we work with scale invariant variables: $ v = V_0 v, ~\pp = V_0 \pp, ~p_T = V_0 p_T$ ($p_v, \phi$ and $T$ are already scale invariant). This ensures that our results do not depend on the unphysical $V_0$.

We now proceed to use the dust field as a clock, explicitly solve the Hamiltonian constraint, and get to a reduced physical Hamiltonian \cite{Husain:2011tk, Husain:2011tm}. We follow the conventions of \cite{Hassan:2017cje}, where we first identify our time variable with the dust field: $t=-T$. The Hamiltonian constraint (\ref{Ham-constr-cosm}) is then (strongly) solved for the momentum conjugate to this choice of time $p_T$, and the \emph{manifestly positive} physical Hamiltonian becomes (this corresponds to the choice $N=-1$, we refer the reader to \cite{Hassan:2017cje} for details),
\be
\label{phys-ham}
H_p = p_T = -H_G-H_{\phi} = \dfrac{3}{4}v p_v^2  - v \Lambda  - \dfrac{\pp^2}{2 v} - v V(\phi).
\ee

Our next step is to polymer quantize the gravitational sector, and obtain an effective background by taking expectation values of the quantum operators in suitably constructed semi-classical gravitational states. On this effective background, we then polymer quantize the scalar field, and obtain a semi-classical Hamiltonian by taking expectation values in semi-classical scalar field states. We proceed to describe the process of polymer quantization in the following subsection.

\subsection{Polymer quantization}\label{Gravity}

Polymer quantization is a method of quantization inspired by LQG, and different from usual Schrodinger quantization. It utilizes a different representation of phase space variables and Hilbert space than that of standard canonical quantization. For the FLRW universe, we take the basic variables to be the volume $v$, and the exponential of the momentum conjugate to the volume $U \equiv \exp(i \lv p_v)$ which satisfy the Poisson bracket relation,
\be
\{v,U\} = i \lv U.
\ee
Polymer quantization proceeds by promoting these variables to operators on the Hilbert space $L^2(\mathbb{R}_{Bohr})$, where $\mathbb{R}_{Bohr}$ is the Bohr compactification of the real line. A basis for this Hilbert space is provided by the states $|\nu\big>$ with the inner product,
\be
\big{<}\nu|\nu'\big> = \delta_{\nu, \nu'},
\ee
where $\delta_{\nu, \nu'}$ is the generalization of the Kronecker delta to real numbers. The action of the basic operators on these states is given as,
\be
\widehat{v}  |\nu \big{>} = \nu |\nu \big{>} ~,~ \widehat{U}  |\nu \big{>}= |\nu + \lv \big{>},
\ee
with the commutator relation $[\widehat{v},\widehat{U}] = \lv \widehat{U}$. Therefore, the states $|\nu\big>$ are eigenstates of the volume operator $\widehat{v}$ with the eigenvalue $\nu$, and $\widehat{U}$ is a translation operator on these states. The operator $\widehat{U}$ is not weakly continuous in the parameter $\lv$, and hence a momentum operator does not exist. This introduces a fundamental discreteness in the theory with the scale set by $\lv$ - the gravitational polymer scale. A momentum operator can be defined indirectly however, by making use of the expansion of the classical variable $\exp(i \lv p_v)$ as follows,
\be
\widehat{p_v}=\frac{1}{2i{\lambda_v}} \left( \widehat{U}-\widehat{U}^\dag \right).
\ee
From this definition of the momentum operator, the gravitational part of the Hamiltonian becomes (we choose a symmetric operator ordering),
\be
\widehat{H_G}=-\frac{3}{4}\widehat{p_v}\widehat{v}\widehat{p_v}+\Lambda \widehat{v}.
\ee
To study the effective dynamics coming from this Hamiltonian, we compute its expectation value in a suitable `semi-classical' state (following \cite{Husain:2006uh, Taveras:2008ke}). The state is defined as,
\bea
\big{|} \psi \big{>} &=& \frac{1}{\mathfrak{N}}\sum_{-\infty}^{+\infty} B_k  \big{|}\nu_k \big{>}, \nn \\
B_k &=& \exp{\left({\dfrac{-(\nu_k-v)^2}{2\sigma_{v}^2}}\right)} \exp{({-i p_v \nu_k})},
\eea
where $(v,p_v)$ are the peaking values for this semi-classical state, $\sv$ is the gravitational state width, and $\mathfrak{N}$ is a normalization constant. The expectation value of the Hamiltonian in this state gives,
\be
\label{grav-poly-ham}
\big{<} \widehat{H}_G \big{>}=\frac{3v}{8 \lambda_v^2} \left[ e^{-\lambda_v^2/\sigma_v^2} \cos\left( 2{\lambda_v} p_v \right)-1 \right] + \Lambda v.
\ee
 
\subsection{Polymer quantized gravity with polymer quantized scalar field}

We now minimally couple a massless scalar field to this effective polymer quantized FLRW background, and then polymer quantize the scalar field. This proceeds in a similar fashion to the gravitational variables. We choose as the basic variables $\Phi \equiv v \phi$  and $W \equiv \exp \left( \dfrac{i \lambda_{\phi} p_\phi}{v} \right)$, where $v$ is the peaking value of the gravitational semi-classical state, and is required here to produce the appropriate density weight \cite{Hossain:2009ru}, and $\lp$ is the scalar field polymer scale.

A basis for the scalar field polymer Hilbert space is given by the states $|\mu\big>$, with $\big{<}\mu|\mu'\big> = \delta_{\mu, \mu'}$, and the operator actions,
\be
\widehat{\Phi}  |\mu \big{>} = \mu |\mu \big{>} ~,~ \widehat{W}  |\mu \big{>}= |\mu + \lp \big{>}.
\ee
Once again, a momentum operator for the scalar field does not exist since the translation operator $\widehat{W}$ is not weakly continuous in $\lp$, however, it can be defined indirectly as,
\be
\widehat{\pp}=\frac{v}{2i{\lambda_v}} \left( \widehat{W}-\widehat{W}^\dag \right).
\ee

The semi-classical state for the scalar field is taken to be,
\bea
|\chi\big> &=& \frac{1}{\tilde{\mathfrak{N}}}\sum_{-\infty}^{+\infty} C_k |\mu_k\big>, \nn \\
C_k &=& \exp \left( {\frac{-(\phi_k-\phi)^2}{ {2\sigma_{\phi}^2}}} \right)  \exp \left( -i p_\phi \phi_k \right)
\eea
where  $(\phi,\pp)$ are the peaking values for the scalar field and the scalar field momentum respectively, $\sp$ is the scalar field state width, and $\tilde{\mathfrak{N}}$ is a normalization factor. The expectation value of the scalar field Hamiltonian $\pp^2/2v$ in this state gives,
\be
\label{hphi-eff}
\big{<} \widehat{H}_{\phi} \big{>} = -\frac{v}{4\lambda_{\phi}^2} \left( e^{ -\lambda_{\phi}^2/v^2 \sigma_\phi^2 } \cos\left(\frac{2\lambda_{\phi}p_\phi}{v}\right) -1 \right).
\ee
Combining this with the effective gravitational Hamiltonian (\ref{grav-poly-ham}) gives us the total physical Hamiltonian (\ref{phys-ham}),
\bea
\label{poly-phys-ham}
H_p &=& -\frac{3v}{8 \lambda_v^2} \left( e^{-\lambda_v^2/\sigma_v^2} \cos\left( 2{\lambda_v} p_v \right)-1 \right) - \Lambda v \nn \\
    &+& \frac{v}{4\lambda_{\phi}^2} \left( e^{ -\lambda_{\phi}^2/v^2 \sigma_\phi^2 } \cos\left(\frac{2\lambda_{\phi}p_\phi}{v}\right) -1 \right).
\eea
In the next section, we look at the dynamics generated by this Hamiltonian.

\section{\label{sec-dyn} Semiclassical dynamics}

We now turn to describe the dynamics generated by this semi-classical Hamiltonian - which includes polymer effects in both the gravitational and the matter degrees of freedom - in the dust time gauge. The equations of motion are (where an overdot indicates a derivative with respect to dust time $t$),
\bea
    \dot{v}&=& \frac{3v}{4\lambda_v} e^{-\lambda_v^2/\sigma_v^2} \sin\Big{(}2{\lambda_v} p_v\Big{)},\\
    \dot{p_v} &=& \frac{3}{8\lambda_{v}^2}  e^{-\lambda_v^2/\sigma_v^2} \cos\Big{(}2{\lambda_v} p_v\Big{)} + \Lambda - \frac{3}{8 \lambda_{v}^2} + \frac{1}{4 \lambda_{\phi}^2} \nn \\
    &-& e^{ -\lambda_{\phi}^2/v^2 \sigma_\phi^2 } \Bigg[ \cos\Big{(}\frac{2\lambda_{\phi}p_\phi}{v}\Big{)} \left( \frac{1}{4\lambda_{\phi}^2} + \frac{1}{2v^2\sigma_{\phi}^2} \right) \nn \\
    &+& \sin\Big{(}\frac{2\lambda_{\phi}p_\phi}{v}\Big{)} \left( \frac{p_{\phi}}{2v\lambda_{\phi}} \right) \Bigg], \label{pveqn} \\
     \dot{\phi} &=& -\frac{1}{2\lambda_{\phi}} e^{ -\lambda_{\phi}^2/v^2 \sigma_\phi^2 } \sin\Big{(}\frac{2\lambda_{\phi}p_\phi}{v}\Big{)},\\
     \dot{p_{\phi}} &=& 0.
\eea
and the Hubble parameter $H_b$ is given as,
\be
\label{hubble}
    H_b = \frac{\dot{a}}{a} =\frac{\dot{v}}{3v} =\frac{1}{4\lambda_v} e^{-\lambda_v^2/\sigma_v^2} \sin\Big{(}2{\lambda_v} p_v\Big{)}.
\ee

We solve these equations numerically for different sets of initial conditions $(v, p_v, \phi, \pp)|_{t=0}$, and for different parameter values ($\lv, \sv, \lp, \sp, \Lambda$). The initial conditions are chosen such that the physical Hamiltonian (\ref{poly-phys-ham}) (and therefore the dust energy density) are manifestly positive. We find that a choice of negative dust energy density can lead to bounces. (See also \cite{Husain:2011tm, Giesel:2020raf} for comments on the role of a negative dust energy density, which leads to unphysical bounces - with negative matter energy densities, a bounce can be achieved even in the classical theory.) We follow the convention of \cite{Hassan:2017cje}, where we only start with initial conditions that give rise to a non-negative dust energy density. And since the physical Hamiltonian in dust time (\ref{poly-phys-ham}) is a constant of the motion (it does not depend explicitly on time), the dust energy density remains non-negative throughout. This way, any bounces are purely due to quantum effects, and not because of negative energies.

Before proceeding to solving these equations numerically, let us first note some interesting features: (i) The scalar field momentum $\pp$ is a constant of the motion in the absence of a potential term, however, its numerical value does affect the background dynamics since it appears explicitly in the equation of motion (\ref{pveqn}) for $p_v$; (ii) The parameters $\sv$ and $\sp$ represent corrections to the dynamics due to the non-zero width of the semi-classical states. In the appropriate limit ($\sv, \sp \rightarrow \infty$) \cite{Taveras:2008ke, Husain:2006uh} these corrections disappear (as can be seen in the equations above, all the exponential pre-factors reduce to unity, and the $1/\sp^2$ term in (\ref{pveqn}) goes to zero); (iii) The parameters $\lv$ and $\lp$ contain the polymer quantization effects of the gravitational and the scalar field sector respectively. There are two relevant limits to consider here: (a) $\lv \rightarrow 0$. This is the limit in which the polymerization effects in the gravity sector vanish, leaving a polymer quantized scalar field living on a classical FLRW background. Indeed, the equations of motion above reproduce this behaviour, and we recover the results obtained in \cite{Hassan:2017cje}. (b) $\lp \rightarrow 0$. This is the limit in which the polymerization effects in the scalar field sector vanish, leaving a standard scalar field living on a polymer quantized (effective) FLRW background. Again, the equations of motion above reproduce these results in the dust-time gauge (we can compare our equations to those obtained in \cite{Saeed:2024gtk}, but without the Hamiltonian constraint since we have fixed a time gauge) and lastly; (iv) The last three terms in the first line of (\ref{pveqn}) show that both the gravity polymer scale $\lv$ and the matter polymer scale $\lp$ contribute to the cosmological constant term (albeit with opposite signs).

\subsection{Polymer cosmology in dust time \label{dust-cosmo}}

\begin{figure}
\begin{center}
\includegraphics[width=\columnwidth]{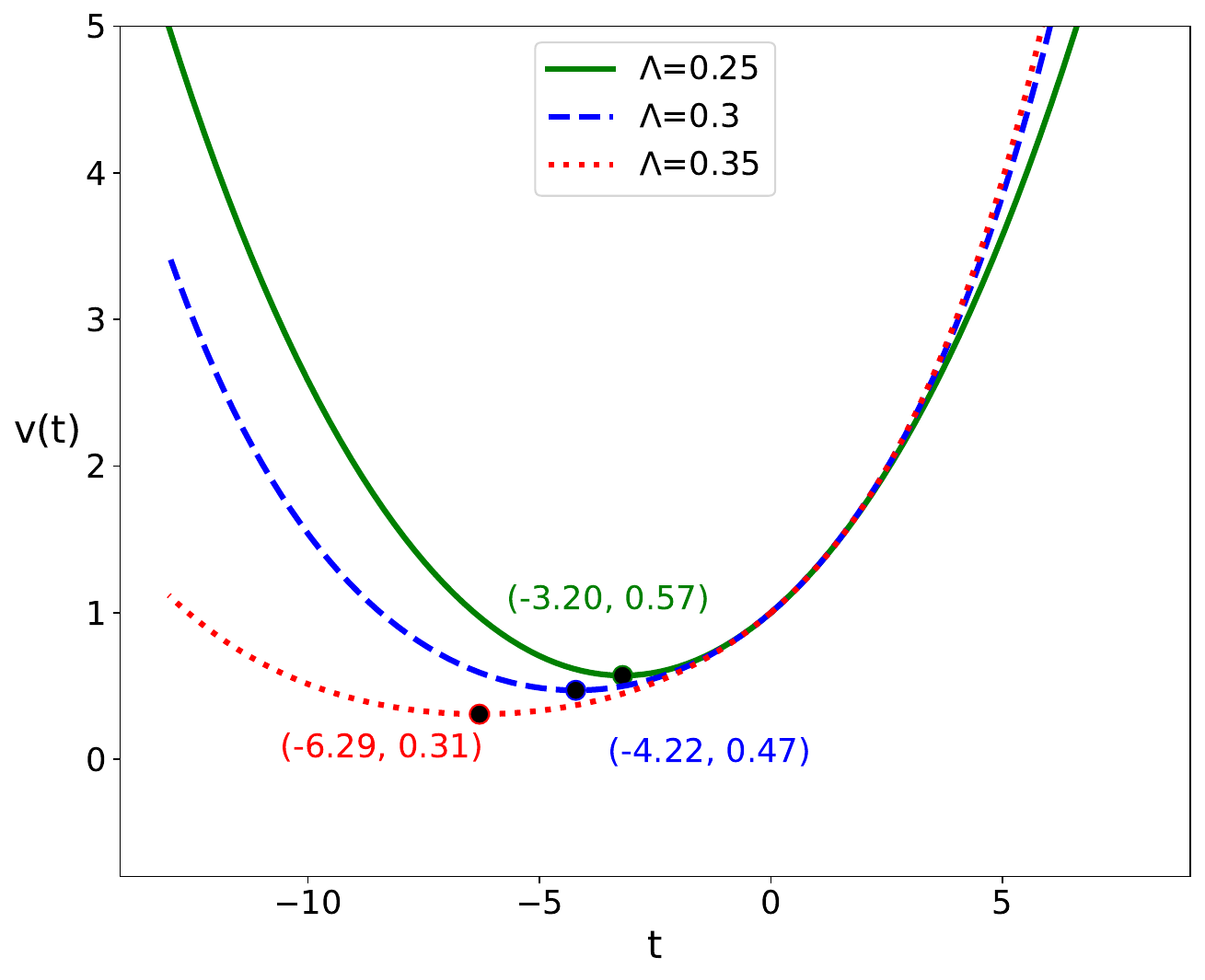}
\end{center}
\caption{\label{fig-gvarylam} Variation in the bounce for the pure gravity case (with dust time) as the cosmological constant $\Lambda$ is varied. The circles show the point of the bounce, and the numerical values indicate the time and the minimum volume at the bounce respectively: $(t_{\text{bounce}}, v_{\text{bounce}})$.}
\end{figure}

\begin{figure}
\begin{center}
\includegraphics[width=\columnwidth]{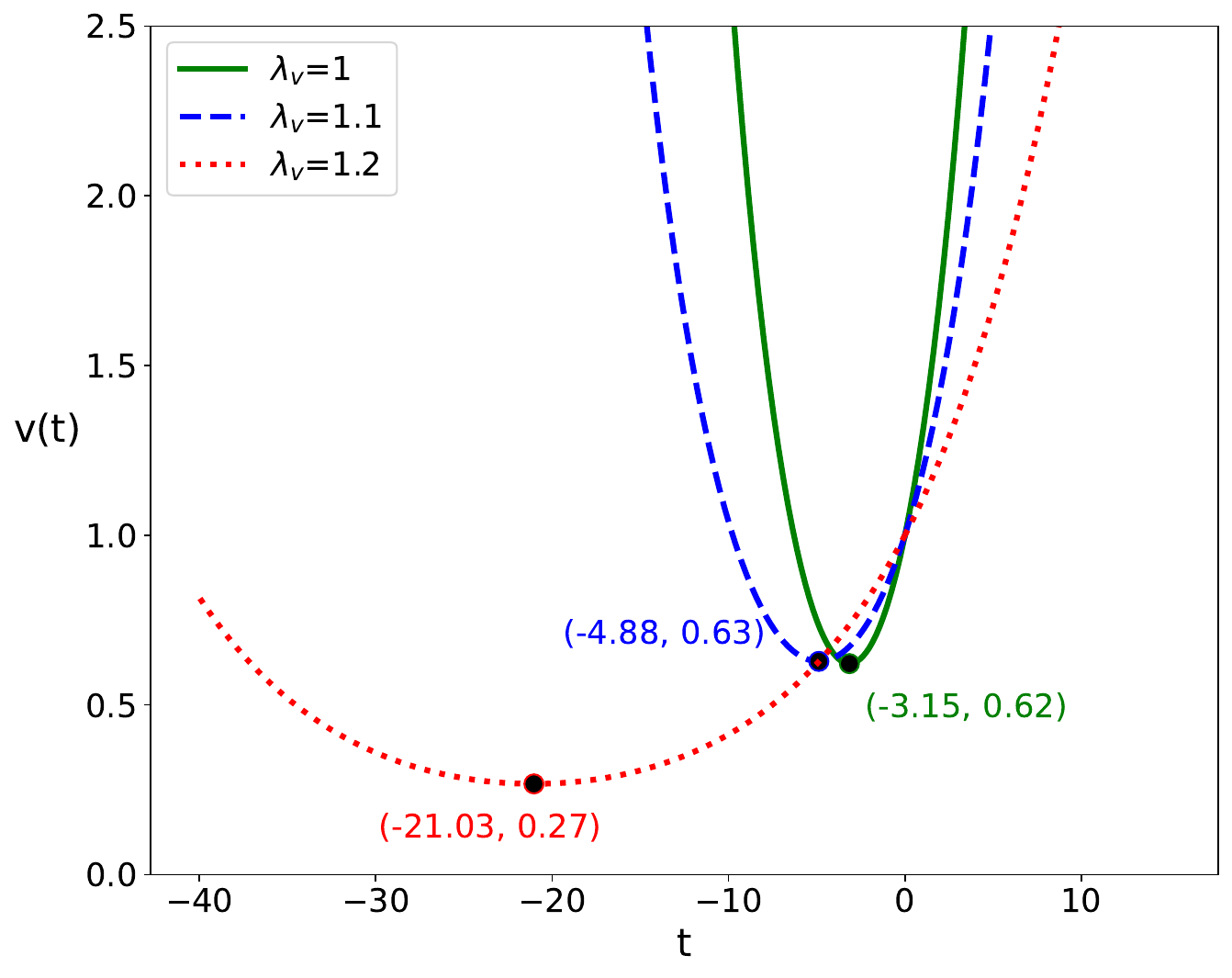}
\end{center}
\caption{\label{fig-gvarylv} Variation in the bounce for the pure gravity case (with dust time) as the gravitational polymer scale $\lv$ is varied. The circles show the point of the bounce, and the numerical values indicate the time and the minimum volume at the bounce respectively: $(t_{\text{bounce}}, v_{\text{bounce}})$.}
\end{figure}

\begin{figure}
\begin{center}
\includegraphics[width=\columnwidth]{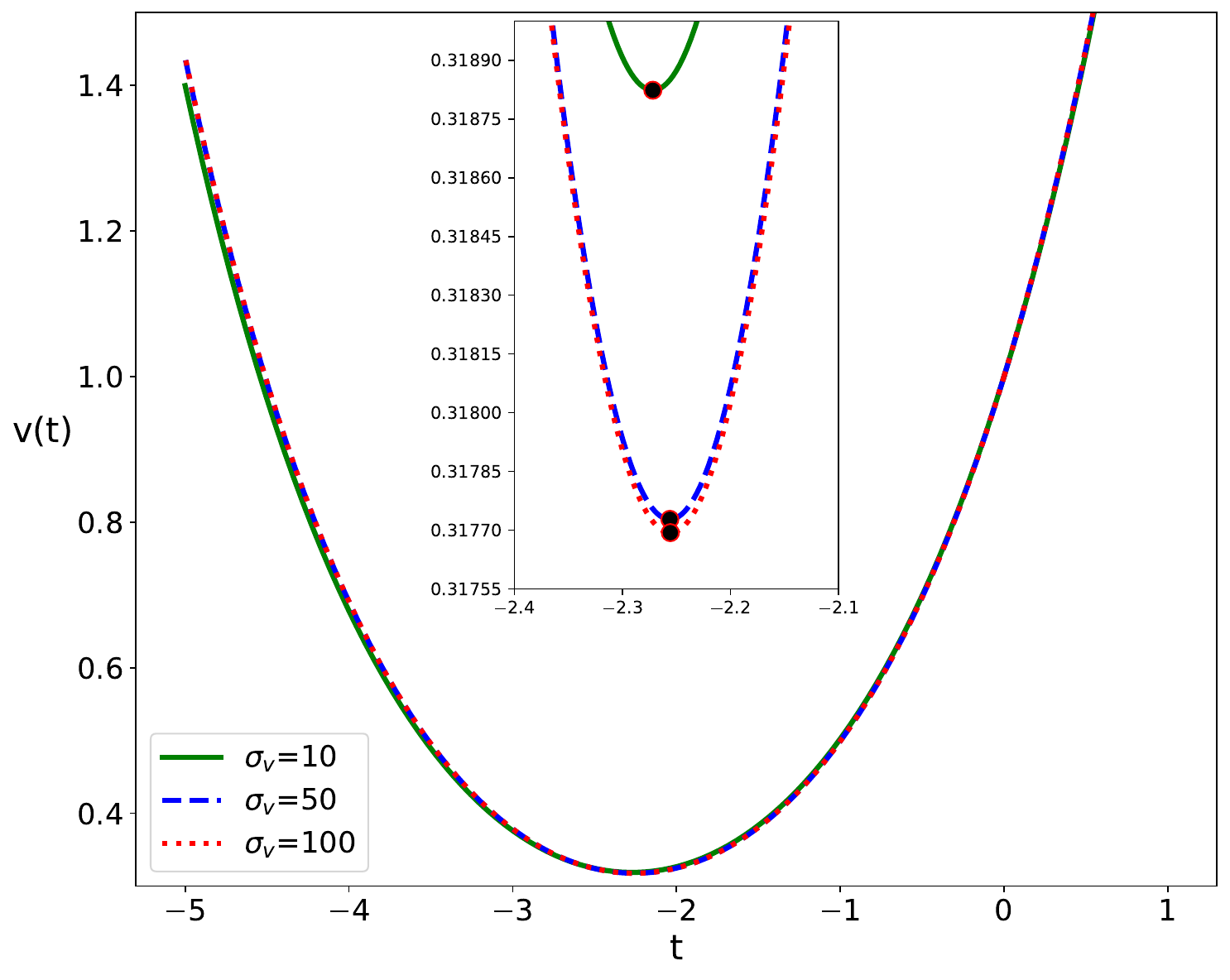}
\end{center}
\caption{\label{fig-gvarysigma} Variation in the bounce for the pure gravity case (with dust time) as the gravitational state width $\sv$ is varied. The inset shows a zoomed in version of the same figure close to the bounce, and the circles indicate the point of the bounce.}
\end{figure}

\begin{figure}
\begin{center}
\includegraphics[width=\columnwidth]{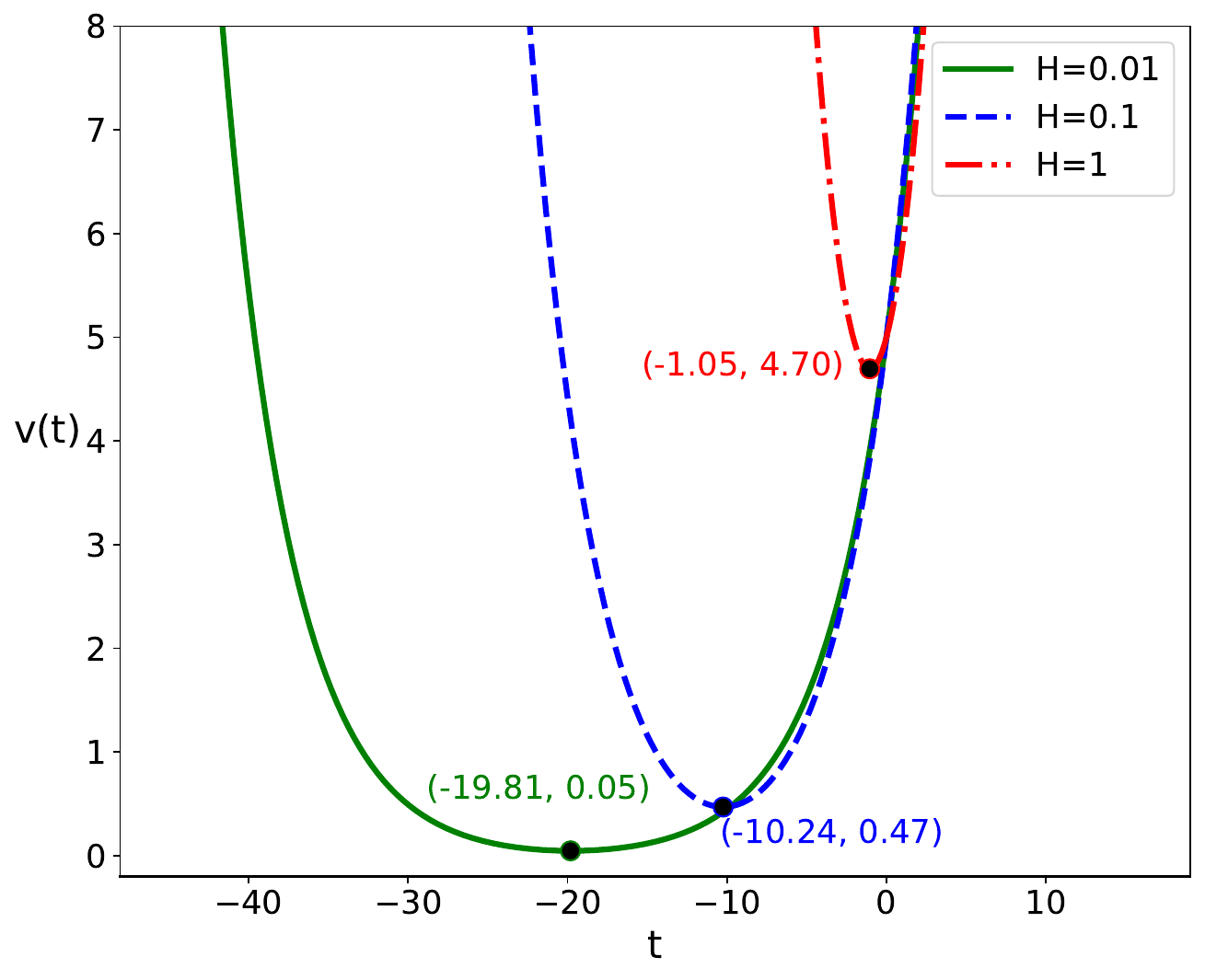}
\end{center}
\caption{\label{fig-gvaryh} Variation in the bounce for the pure gravity case (with dust time) as the value of the total Hamiltonian $H$ is varied. The circles show the point of the bounce, and the numerical values indicate the time and the minimum volume at the bounce respectively: $(t_{\text{bounce}}, v_{\text{bounce}})$. This shows dependence on initial conditions.}
\end{figure}

We first turn to describe the effective dynamics of a polymer quantized FLRW universe in dust time (without the presence of a scalar field). This will set the stage for the subsequent subsection where we include the polymer quantized scalar field. We focus, in particular, on what happens to the quantum bounce as various parameters of interest are changed. Our results are shown in Figures (\ref{fig-gvarylam} - \ref{fig-gvaryh}).

Figure \ref{fig-gvarylam} shows how the time and location of the bounce changes as the value of the cosmological constant $\Lambda$ is changed. The other parameters and initial conditions are kept fixed for these runs at $\lv = 1, \sv = 1, v|_{t=0} = 1, H|_{t=0} = 0.1$. Figure \ref{fig-gvarylv} shows variation in the bounce as the gravitational polymer scale $\lv$ is changed. Other parameters and initial conditions were fixed at $\Lambda=0.3, \sv = 1, v|_{t=0} = 1, p_v|_{t=0} = 1$.

Figure \ref{fig-gvarysigma} shows how the time and location of the bounce changes as the gravitational semi-classical state width parameter $\sv$ is changed. The other parameters and initial conditions are kept fixed for these runs at $\Lambda=0.3, \lv = 1, v|_{t=0} = 1, H|_{t=0} = 0.1$. There is very little change as the width is varied, and it becomes almost insignificant as the width is increased further (the limit in which the semi-classical state is `sharply peaked' corresponds to $\sv \rightarrow \infty$). Figure \ref{fig-gvaryh} shows variation in the bounce as the value of the physical Hamiltonian $H$ is changed (equivalently, this amounts to changing the initial gravitational momentum $p_v$). Other parameters and initial conditions were fixed at $\Lambda=0.3, \lv = 1, \sv = 1, v|_{t=0} = 5$. This illustrates the dependence of the bounce on initial conditions.

\subsection{Polymer cosmology with a polymer scalar field}

\begin{figure}
\begin{center}
\includegraphics[width=\columnwidth]{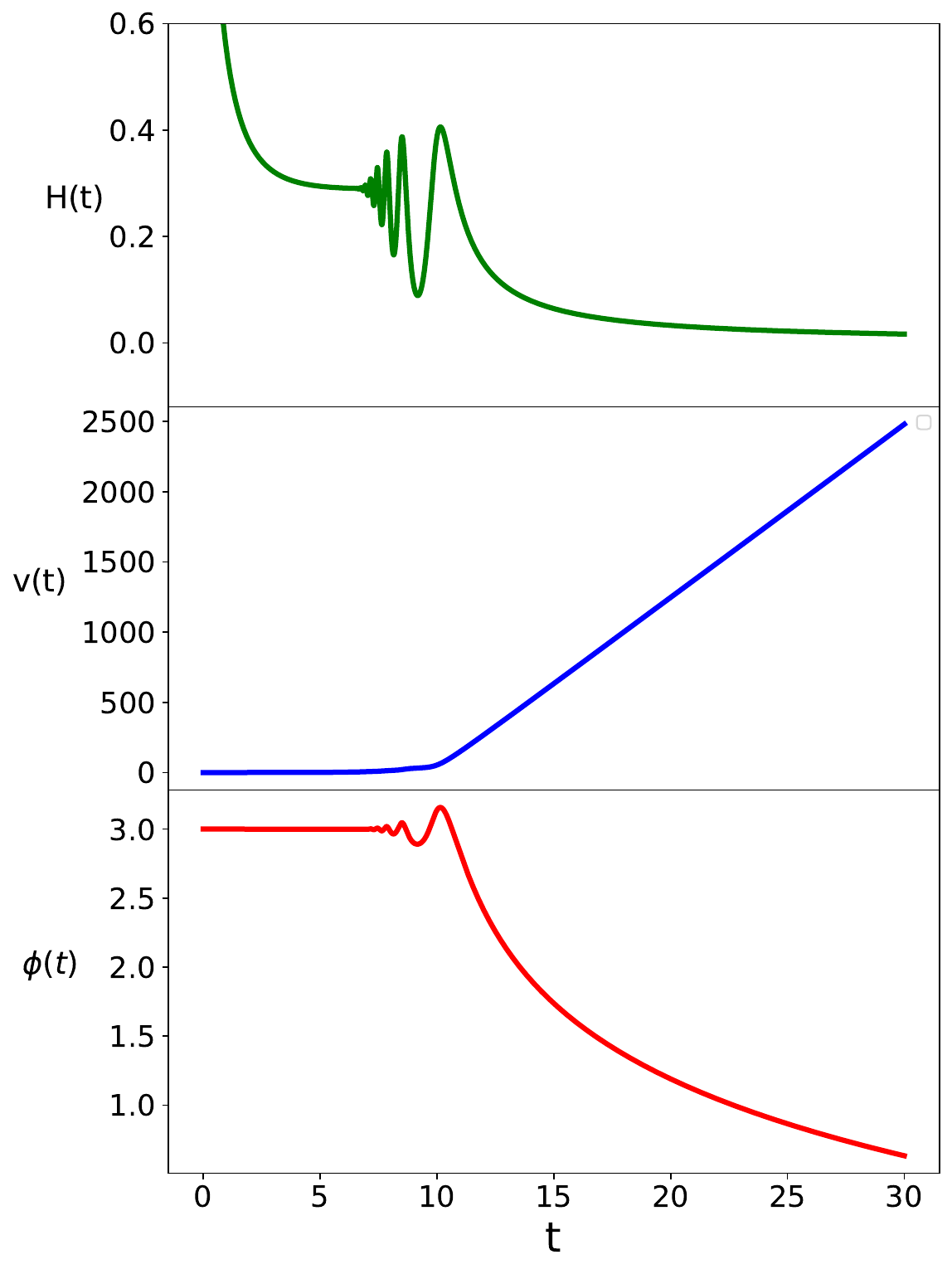}
\end{center}
\caption{\label{fig-lvlimit} The (dust) time evolution of the Hubble parameter, volume and the scalar field for $\lv = 10^{-6}$. In this limit, we have a polymer quantized scalar field living on a classical FLRW background. The results show that there is no bounce and hence the early time singularity is present.}
\end{figure}

\begin{figure}
\begin{center}
\includegraphics[width=\columnwidth]{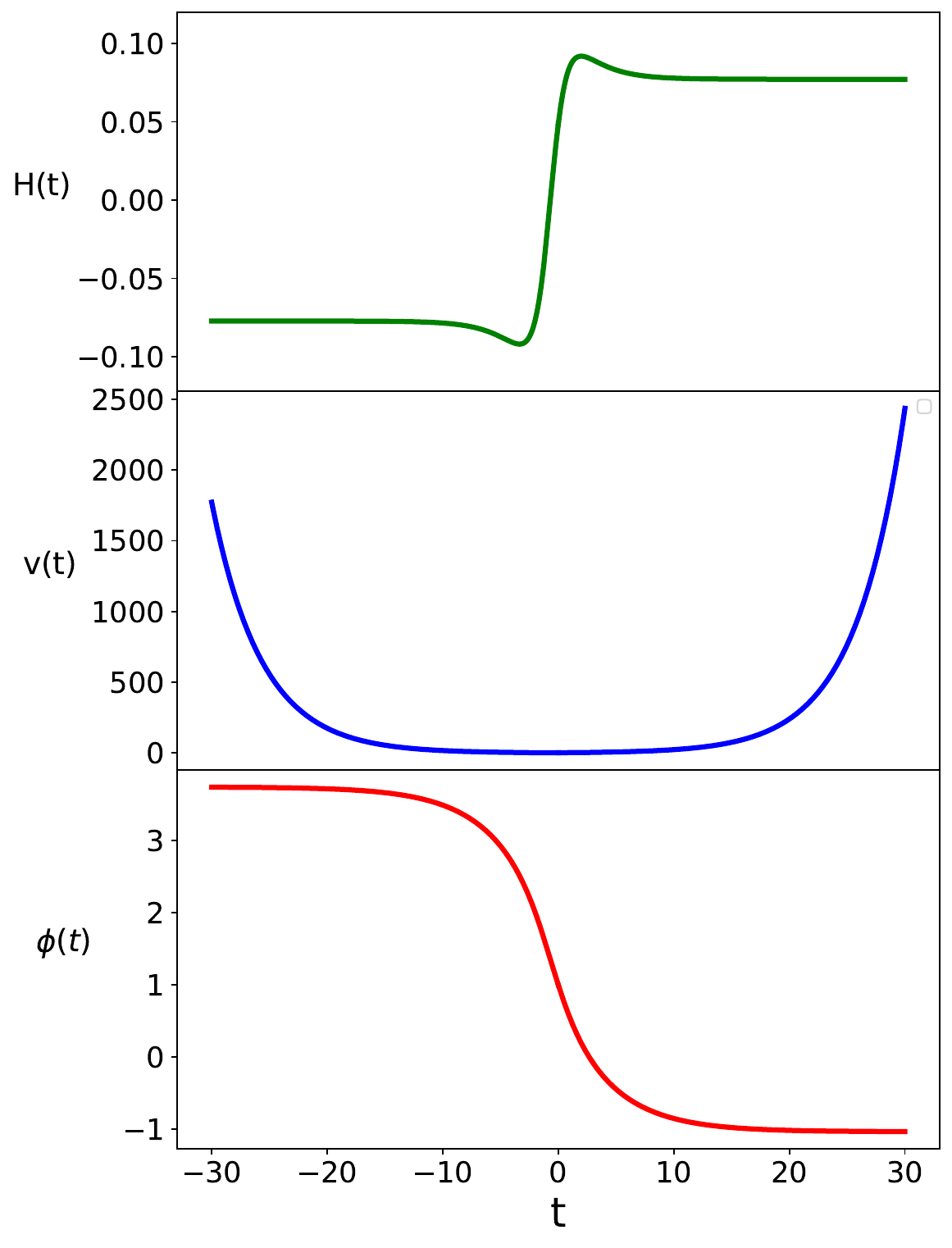}
\end{center}
\caption{\label{fig-lplimit} The (dust) time evolution of the Hubble parameter, volume and the scalar field for $\lp = 10^{-6}$. In this limit, we have a standard scalar field living on an effective polymer quantized FLRW background. The results show a bounce in the early universe.}
\end{figure}

\begin{figure}
\begin{center}
\includegraphics[width=\columnwidth]{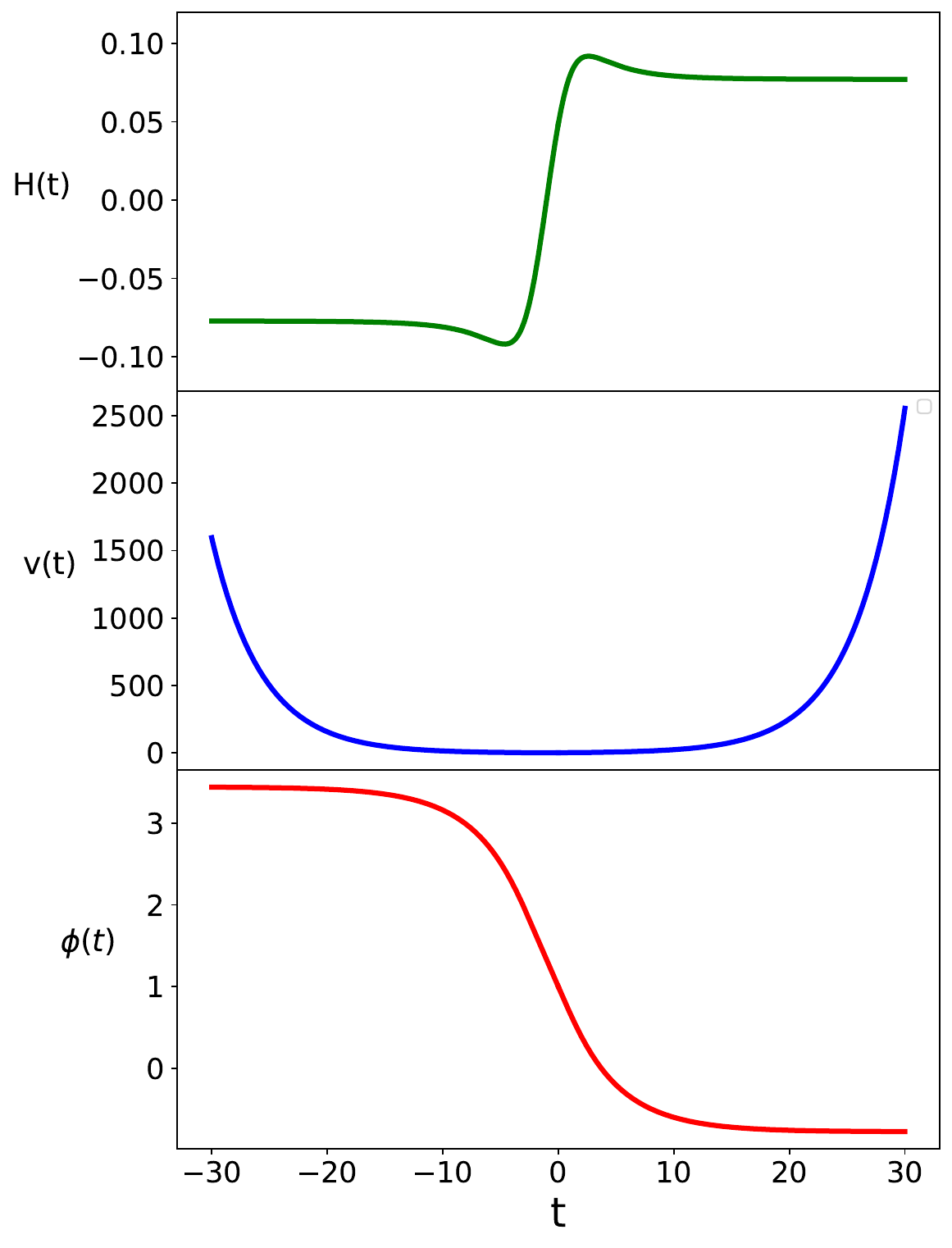}
\end{center}
\caption{\label{fig-dynamicslvlp} Time evolution of the Hubble parameter, volume and the scalar field for $\lv = 1, \lp = 1, \sv = 1 = \sp, \Lambda = 0.3$. There is a bounce, with expansion taking place before and after the bounce.}
\end{figure}

\begin{figure}
\begin{center}
\includegraphics[width=\columnwidth]{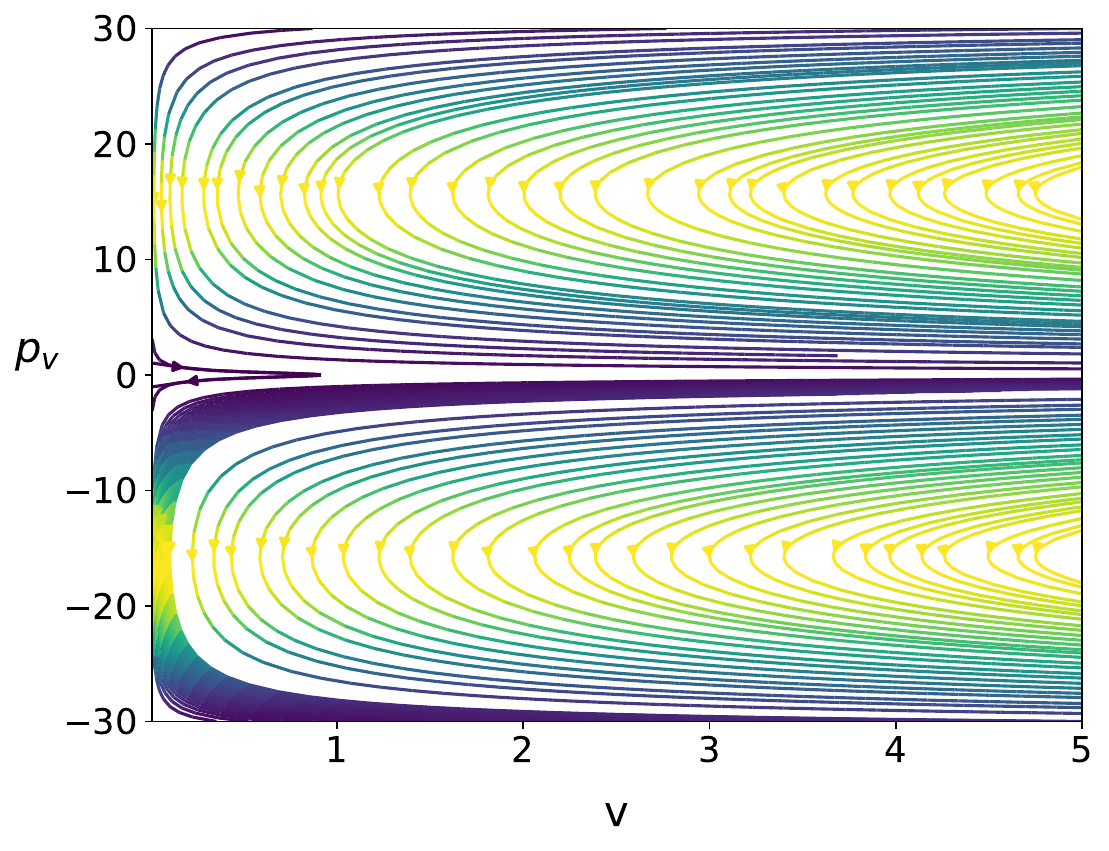}
\end{center}
\caption{\label{fig-phase1} Phase portrait of $(v,p_v)$. There is clear dependence of the bounce on initial conditions.}
\end{figure}

\begin{figure}
\begin{center}
\includegraphics[width=\columnwidth]{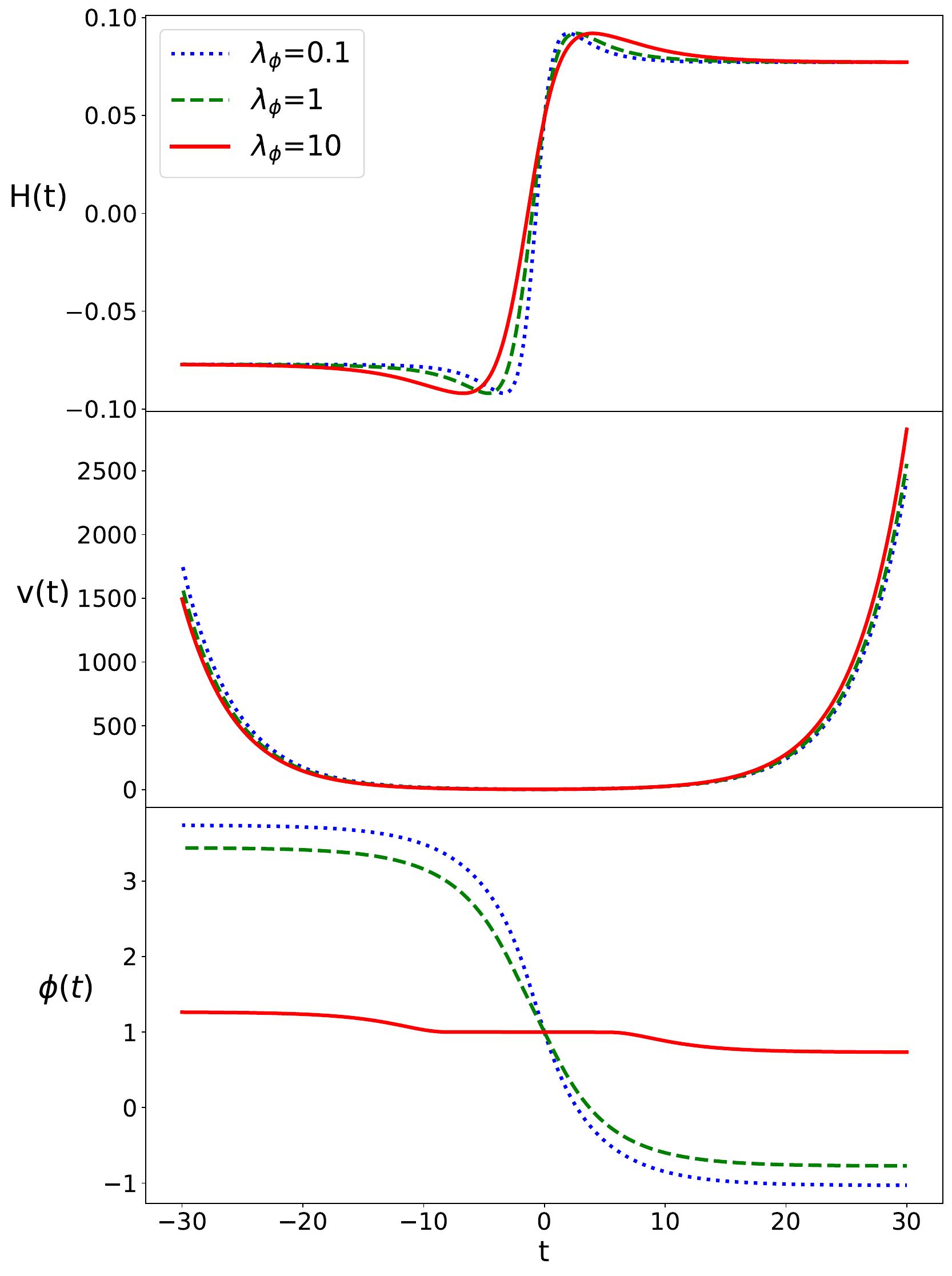}
\end{center}
\caption{\label{fig-dynamicsvarylp} Variation in the evolution of the Hubble parameter, volume and scalar field as the scalar field polymer scale $\lp$ is varied.}
\end{figure}

\begin{figure}
\begin{center}
\includegraphics[width=\columnwidth]{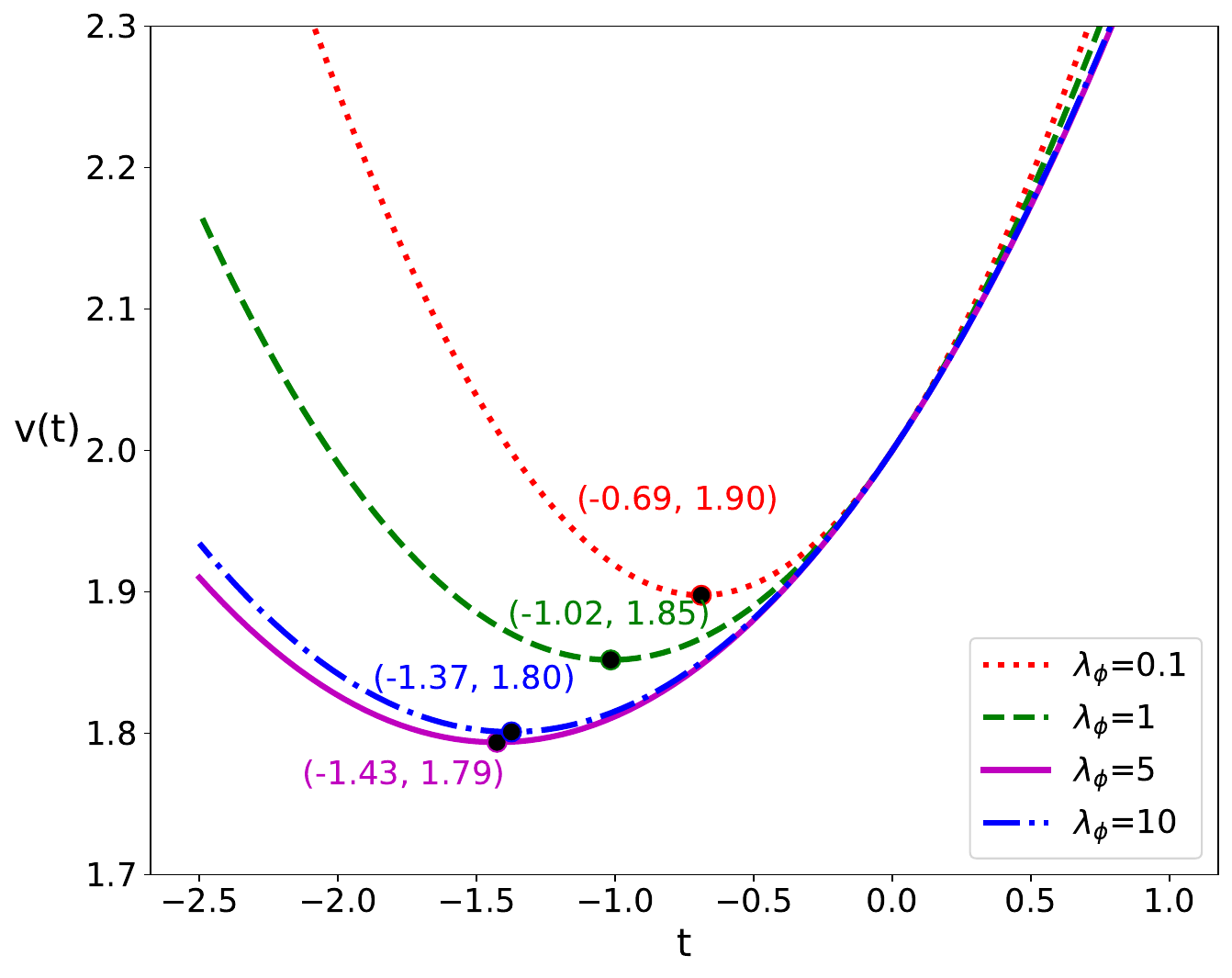}
\end{center}
\caption{\label{fig-bouncevarylp} Variation in the location of the bounce as the scalar field polymer scale $\lp$ is varied. The circles show the point of the bounce, and the numerical values indicate the time and the minimum volume at the bounce respectively: $(t_{\text{bounce}}, v_{\text{bounce}})$.}
\end{figure}

\begin{figure}
\begin{center}
\includegraphics[width=\columnwidth]{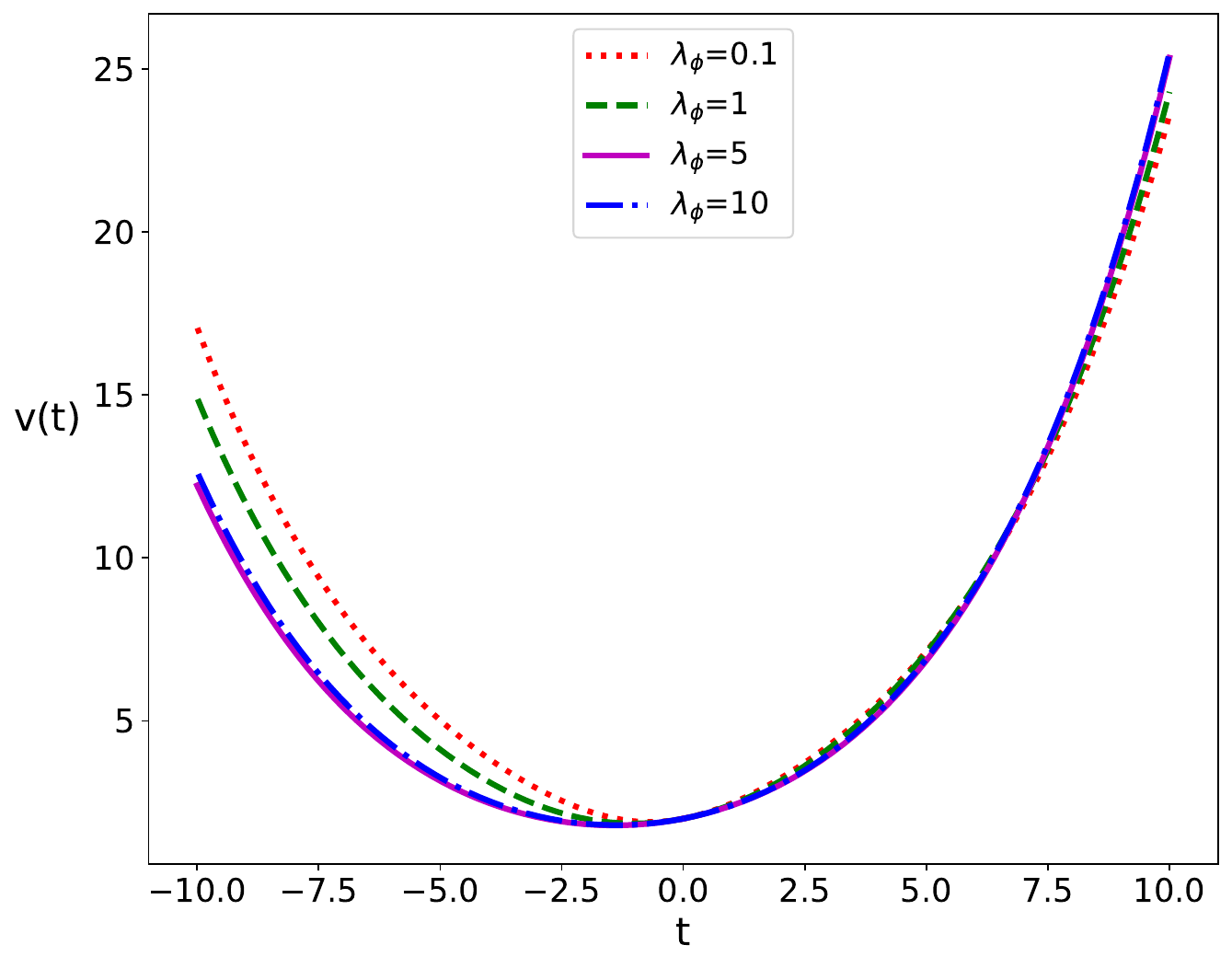}
\end{center}
\caption{\label{fig-bouncevarylp-large} Variation in the dynamics of the volume of the universe as the scalar field polymer scale $\lp$ is varied. There is asymmetric evolution across the bounce.}
\end{figure}

\begin{figure}
\begin{center}
\includegraphics[width=\columnwidth]{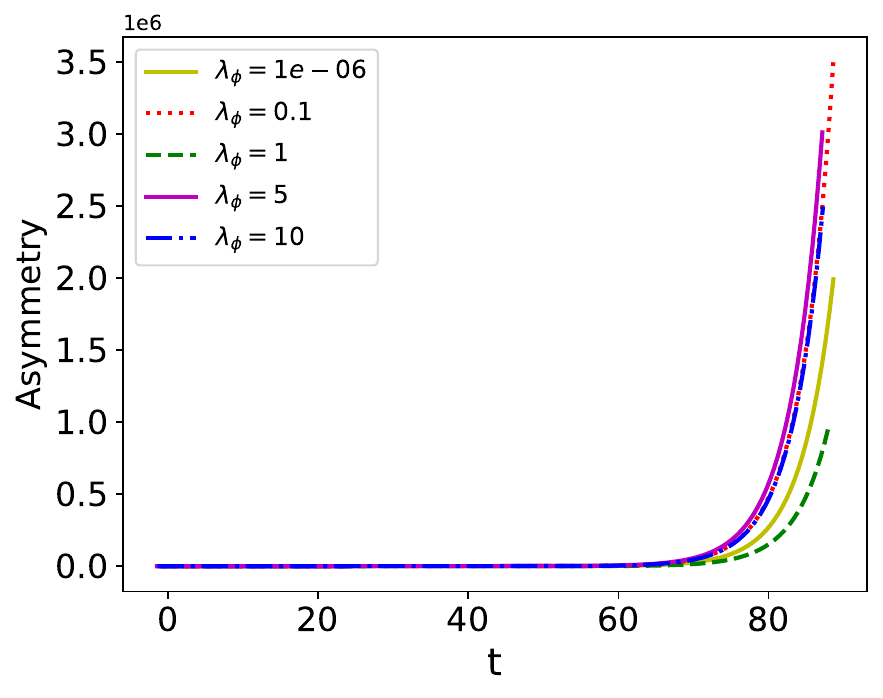}
\end{center}
\caption{\label{fig-asym} Asymmetry vs time. Asymmetry is computed by taking the absolute value of the difference between the volume after the bounce, and the volume before the bounce. Different values of the matter polymer scale produce different asymmetries, and the asymmetry increases as time goes on.}
\end{figure}

\begin{figure}
\begin{center}
\includegraphics[width=\columnwidth]{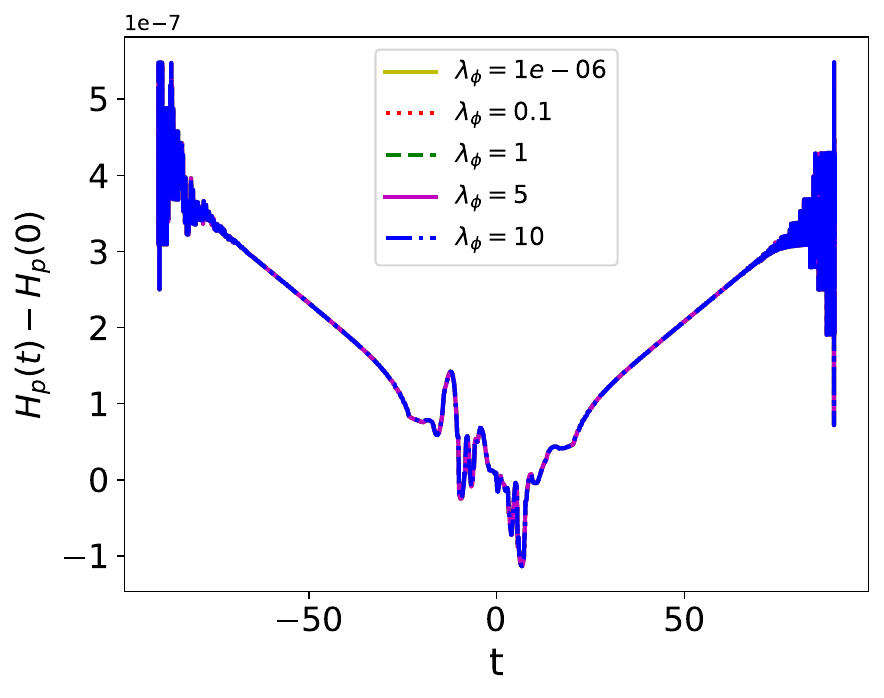}
\end{center}
\caption{\label{fig-ham} The difference of the physical Hamiltonian $H_p$ from its initial value $H_p(0)$ as a function of time. There is very minor variation over a substantial time range indicating good control over the numerics.}
\end{figure}

We now solve the full set of equations (with the polymer quantized scalar field included) numerically. Our results appear in Figures (\ref{fig-lvlimit} - \ref{fig-ham}). There are two relevant limits to consider here. First is the limit $\lv \rightarrow 0$. This is the case when the gravitational polymer scale vanishes, and we are left with a polymer quantized scalar field living on a classical background. Figure \ref{fig-lvlimit} shows the Hubble parameter, volume and scalar field for this case. As expected, there is a singularity at early times since the gravitational sector is classical. This limit reproduces the results of \cite{Hassan:2017cje}. The other limit is the case when the scalar field polymer scale vanishes: $\lp \rightarrow 0$. This produces an ordinary scalar field living on an effective polymer quantized background. The results are shown in Figure \ref{fig-lplimit}, where we can see the Hubble parameter, volume and scalar field. The early time singularity is resolved.

Beyond these limits, that serve as a useful check of our model, we are interested in what happens under arbitrary values of these parameters. Figure \ref{fig-dynamicslvlp} shows the results of a generic evolution. The initial conditions were $(v, p_v, \phi, \pp)|_{t=0} = (2, 1.29, 1, 1)$. There is a bounce at early times, followed by expansion before and after the bounce. To see what effect varying the initial conditions has, we constructed a phase portrait for the gravitational variables $(v,p_v)$ shown in Figure \ref{fig-phase1} (since the scalar field momentum $\pp$ is a constant of the motion, the phase portrait of $(\phi, \pp)$ is trivial, and we omit it here). The various parameter values were set at $\Lambda = 0.3, \lv = 0.01, \lp = 1, \sv = 1 = \sp$, and the initial conditions for the scalar field were $(\phi, \pp)|_{t=0} = (1, 1)$. We see that a bounce happens for all of these initial conditions (i.e., it is a generic feature of the theory and is robust under changes in initial conditions), however, the location of the bounce depends on the initial conditions.

In Sec \ref{dust-cosmo}, we looked at what happens to the evolution of the volume (in particular, what happens to the bounce) in the absence of the scalar field, as we vary parameters such as the cosmological constant $\Lambda$, the gravitational polymer scale $\lv$, and the gravitational state width $\sv$. Now, with the (polymer-quantized) scalar field added, we can again look at the effect of varying these parameters, however, the results are qualitatively similar. Instead, we focus on the new physical parameter - the scalar field polymer scale $\lp$ - and see how changing it affects the evolution.

Figure \ref{fig-dynamicsvarylp} shows the time evolution of the Hubble parameter, volume and the scalar field for various values of $\lp$. Other parameters and initial conditions were kept fixed at $ \Lambda = 0.3, \lv = 1, \sv = 1, \sp = 1 $ and $(v, p_v, \phi, \pp)|_{t=0} = (2, 1.29, 1, 1)$ respectively. Figure \ref{fig-bouncevarylp} shows a closeup of the volume $v(t)$ to illustrate how the bounce changes as $\lp$ is varied for the same other parameter values and initial conditions. There is a significant change in both the time and the location of the bounce as $\lp$ is varied. This shows that it is not only the gravitational parameters, but also the polymer scale of the matter sector that affects evolution in the early universe.

Lastly, we note that, in this model there is asymmetric evolution across the bounce. Figure \ref{fig-bouncevarylp-large} shows the same evolution as Figure \ref{fig-bouncevarylp}, but on a slightly larger timescale so that the asymmetry becomes evident. In Figure \ref{fig-asym}, the asymmetry (which is computed by taking the absolute value of the difference between the volume after the bounce, and the volume before the bounce) is shown for a longer time evolution making the difference clear. Finally, Figure \ref{fig-ham} shows the time evolution of the physical Hamiltonian (after subtracting the initial value), serving as a useful numerical check of our code (since the Hamiltonian must remain constant). We find that, over a large range of parameter values, the numerics remain accurate.

We have focused here on a discussion of the effects of polymer quantization of both the matter and the gravitational sector (jn particular the effects of a matter polymer scale on the quantum bounce) in the dust time gauge. To see how this model connects to the usual discussion in the literature in terms of matter energy densities, and the effective Friedmann equation, we refer the reader to the Appendix (\ref{Appx}).

\section{\label{sec-con} Summary}

We studied the semi-classical dynamics of a polymer quantized scalar field living on an effective polymer quantized FLRW background using a pressureless dust field as time. While polymer quantization effects in the gravitational sector, and in the matter sector have been studied before separately, here we presented a model that includes these effects - at the semi-classical level - in both of these sectors. From a fundamental perspective, if gravity is indeed polymer (or loop) quantized, then it is natural to expect that the matter sector should also be polymer quantized.

In the purely gravitational case, with dust as time, we found that there is a bounce at early times, as expected from a polymer (or loop) quantization of an FLRW universe. Furthermore, we found that the location of this bounce (both the time at which it occurs, and the minimum value that the volume takes) changes if we change the gravitational polymer scale $\lv$, or the gravitational state width $\sv$, or the value of the cosmological constant $\Lambda$ for the \emph{same} initial conditions. The bounce also depends on the initial conditions, and changes as the value of the total physical Hamiltonian is changed.

With the addition of a polymer quantized scalar field, we note that there is a significant effect on the dynamics, and in particular, the location and time of the bounce changes as the scalar field polymer scale $\lp$ is varied. There is also dependence on initial conditions, and the evolution across the bounce is asymmetric.

On a broader scale, this project investigates the effects of quantum gravity corrections on both the gravitational and the matter sectors. While we have restricted ourselves here to a semi-classical treatment, this project serves as a prelude to a fuller treatment of polymer quantized matter coupled to a polymer quantized background with full backreaction included. Indeed, standard quantum field theory of matter fields on a quantized gravitational background has been studied before \cite{Ashtekar:2009mb}. It will be interesting to note what effects, if any, are introduced by the polymer quantization of matter.

\begin{acknowledgments}
SMH would like to thank Ammar Ahmed Khan for helpful discussions. This project was supported by the Higher Education Commission (HEC) of Pakistan through NRPU grant No. 20-15435.
\end{acknowledgments}

\bibliography{poly-cosmo-matter}

\appendix
\section{\label{Appx} The effective Friedmann equation}

It is standard in the literature to describe the dynamics of the bounce in terms of matter energy densities. In this appendix, we provide that description for our model. We have relegated this discussion to an appendix as our main emphasis is different, for two reasons: (i) We work in the dust time gauge after making an explicit gauge fixing at the classical level in the theory. This means that the Hamiltonian constraint has been (strongly) solved, and no longer appears as a constraint; and (ii) We want to elucidate the effects of the polymer matter scale ($\lp$) on the quantum gravitational bounce which would not be possible if matter parameters remain `hidden' in the energy densities. Nevertheless, to connect with discussion in the literature, and to show that our model also reproduces and extends the results reported before, we provide this description here.

To begin, we have to rewrite the Hamiltonian constraint ($\mathcal{H} = - H_P + P_T \approx 0$) as the Friedmann equation (in terms of the Hubble parameter $H_b$, and the matter energy densities). The Hubble parameter is given in (\ref{hubble}), the energy density corresponding to the cosmological constant is $\rho_{\Lambda} = \Lambda$, the energy density of the dust field is $\rho_D = P_T/v$ (since $P_T$ is the dust Hamiltonian density), and the scalar field energy density is $\big{<} \widehat{\rho}_{\phi} \big{>} = \big{<} \widehat{H}_{\phi} \big{>}/v$ where $\big{<} \widehat{H}_{\phi} \big{>}$ is given in (\ref{hphi-eff}). This gives the total matter energy density as, $\rho_\text{tot} = \big{<} \widehat{\rho}_{\phi} \big{>} + \rho_D + \rho_{\Lambda}$.

Solving the Hamiltonian constraint in terms of these variables then gives us the effective Friedmann equation,
\be
H_b^2 = \frac{\rho_\text{tot}}{3} \left( 1 - \frac{\rho_\text{tot}}{\rho_0} \right) + f(\lv, \sv),
\ee
where,
\be
\rho_0 = \frac{3}{4 \lv^2},
\ee
and,
\be
f(\lv, \sv) = \dfrac{e^{-2 \lv^2/ \sv^2} - 1}{16 \lv^2}
\ee
is a function that encodes the effects of a finite polymer scale and a finite state width.

The bounce occurs when the Hubble parameter crosses zero. Solving for the critical density $\rho_c$ at which this occurs gives,
\be
\rho_c = \frac{3}{8 \lv^2} \left( 1 + e^{-\lv^2/ \sv^2} \right),
\ee
which, in the limit of vanishing polymer scale ($\lv \rightarrow 0$) or infinite state width ($\sv \rightarrow \infty$), reduces to $\rho_0$ above.

Given this critical energy density, we can also solve for the volume at which the bounce occurs ($v_{\text{bounce}}$) in terms of various matter and gravitational parameters. However, in this case, it leads to a transcendental equation, and has to be solved numerically. In the special limit of infinite state widths ($\sv \rightarrow \infty, \sp \rightarrow \infty$), and vanishing dust field ($\mathcal{H}_D \rightarrow 0$), we can recover an analytic solution,
\be
v_{\text{bounce}} = \frac{2 \lp P_\phi}{\cos^{-1} \left( 1 + 4 \lp^2 \Lambda - 3 \dfrac{\lp^2}{\lv^2} \right)}.
\ee
In the limit of vanishing cosmological constant ($\Lambda \rightarrow 0$), and vanishing matter polymer scale ($\lp \rightarrow 0$), this reproduces the result found in \cite{Saeed:2024gtk}.

\end{document}